\documentclass[11pt]{article}
\usepackage[top = 2 cm, bottom = 2 cm, left = 2 cm, right = 1.5 cm]{geometry}
\usepackage{authblk, cite, color, amssymb, amsmath}
\usepackage[hypertex, colorlinks = true, linkcolor = blue, citecolor = red]{hyperref}
\usepackage[dvipsnames]{xcolor}

\begin{document}

\title{Geometry of the Non-Compact G(2)}

\author{{\bf Merab Gogberashvili$^{1,2}$} and {\bf Alexandre Gurchumelia}$^{1}$}
\affil{\small $^1$ Javakhishvili Tbilisi State University, 3 Chavchavadze Avenue, Tbilisi 0179, Georgia \authorcr
$^2$ Andronikashvili Institute of Physics, 6 Tamarashvili Street, Tbilisi 0177, Georgia}
\maketitle

\begin{abstract}

Geometrical applications of the non-compact form of Cartan's exceptional Lie group G(2) is considered. This group generates specific rotations of 7-dimensional Minkowski-like space with three extra time-like coordinates and in some limiting cases imitates standard Poincare transformations. In this model space-time translations are non-commutative and are represented by the rotations towards the extra time-like coordinates. The second order Casimir element of non-compact G(2) and its expression by the Casimir operators of the Lorentz and Poincare groups are found.


\end{abstract}


In our previous papers geometrical applications of split octonions over the field of real numbers was considered \cite{Gogberashvili:2015sga, Gogberashvili:2016ztr, Gogberashvili:2005xb, Gogberashvili:2004na, Gogberashvili:2002wf}. Space-time symmetries in this model are represented by the 14-parameter automorphism group of split octonions, $G^{NC}_2$. Generators of $G_2^{NC}$ were first provided by \'{E}lie Cartan in the following form \cite{Cart}:
\begin{equation} \label{CartanTransformations}
\begin{split}
Y_{kk} &= -z_k\frac{\partial}{\partial z^k} + y_k\frac{\partial}{\partial y^k} + \frac{1}{3} \sum_i \left(z_i\frac{\partial}{\partial z_i}-y_i\frac{\partial}{\partial y_i}\right)~,\\
Y_{k0} &= -2t\frac{\partial}{\partial z^k} + y_k\frac{\partial}{\partial t} + \frac{1}{2} \sum_{i,j}\epsilon_{ijk}\left(z^i\frac{\partial}{\partial y_j}-z^j\frac{\partial}{\partial y_i}\right)~,\\
Y_{0k} &= -2t\frac{\partial}{\partial y^k} + z_k\frac{\partial}{\partial t} + \frac{1}{2} \sum_{i,j}\epsilon_{ijk}\left(y^i\frac{\partial}{\partial z_j}-y^j\frac{\partial}{\partial z_i}\right)~,\\
Y_{ij} &= -z_j\frac{\partial}{\partial z^i}+y_i\frac{\partial}{\partial y^j}~. \qquad \qquad \qquad (i,j,k = 1,2,3)
\end{split}
\end{equation}
Due to the constraint,
\begin{equation}
\label{constraint}
Y_{11} + Y_{22} + Y_{33} = 0 ~,
\end{equation}
only 14 of 15 operators (\ref{CartanTransformations}) are linearly independent.

One can write the generators (\ref{CartanTransformations}) as the $7 \times 7$-matrices,
\begin{equation} \label{Matrix}
\begin{pmatrix}
\textbf{A}(\alpha) & 2d & \textbf{B}(b)\\
-b^T & 0 & -d^T \\
\textbf{B}(d) & 2b & -\textbf{A}^T(\alpha)
\end{pmatrix}=
\alpha_{0k}Y^{0k} + \alpha_{k0}Y^{k0} + \alpha_{ij}Y^{ij}  ~,
\end{equation}
which act on a 7 dimensional vector
\begin{equation} \label{q}
p =
\begin{pmatrix}
y_k\\
t\\
z_k
\end{pmatrix} ~. \qquad \qquad (k = 1,2,3)
\end{equation}
In (\ref{Matrix}) $\alpha_{\mu\nu}$ ($\mu, \nu = 0,1,2,3$) are group parameters, the quantities $b$ and $d$ denote column vectors,
\begin{equation}
b = \begin{pmatrix}
\alpha_{10}&\alpha_{20}&\alpha_{30}
\end{pmatrix}^T~, \qquad \qquad
d = \begin{pmatrix}
\alpha_{01}&\alpha_{02}&\alpha_{03}
\end{pmatrix}^T~,
\end{equation}
the $3\times 3$-matrix $\textbf{A}$ is a $SU(3)$ group generator,
\begin{equation} \label{A}
\textbf{A}(\alpha) = \frac{1}{3}\begin{pmatrix}
-2\alpha_{11}+\alpha_{22}+\alpha_{33}&-3\alpha_{21}&-3\alpha_{31}\\
-3\alpha_{12}&\alpha_{11}-2\alpha_{22}+\alpha_{33}&-3\alpha_{32}\\
-3\alpha_{13}&-3\alpha_{23}&\alpha_{11}+\alpha_{22}-2\alpha_{33}
\end{pmatrix}~,
\end{equation}
and the two $3\times 3$-matrices $\textbf{B}$ are $SO(3)$ generators,
\begin{equation} \label{B}
\textbf{B}(b) = \begin{pmatrix}
0&\alpha_{30}&-\alpha_{20}\\
-\alpha_{30}&0&\alpha_{10}\\
\alpha_{20}&-\alpha_{10}&0
\end{pmatrix}~, \qquad \qquad
\textbf{B}(d) = \begin{pmatrix}
0&\alpha_{03}&-\alpha_{02}\\
-\alpha_{03}&0&\alpha_{01}\\
\alpha_{02}&-\alpha_{01}&0
\end{pmatrix}~.
\end{equation}

In the representation (\ref{CartanTransformations}) the invariant quadratic form, which is conserved under the $G_2^{NC}$ transformations has the form:
\begin{equation} \label{metric}
p^T \texttt{g} \,p = t^2 + z_ky^k~,
\end{equation}
where
\begin{equation}
\texttt{g} = \frac{1}{2}\begin{pmatrix}
0&0&\textbf{1}\\
0&2&0\\
\textbf{1}&0&0
\end{pmatrix}~,
\end{equation}
is the $7\times7$ metric tensor and $\textbf{1}$ denotes the $3 \times 3$ identity matrix.

We want to give geometrical interpretation to the invariant quadratic form of the $G_2^{NC}$ group and it is convenient to transform (\ref{metric}) in the Minkowski-like diagonal form. For this purpose let us perform the similarity transformation of the generators (\ref{Matrix}),
\begin{equation}
X = \Omega \,Y \Omega^{-1}~,
\end{equation}
where
\begin{equation}
\Omega = \frac{1}{2}\begin{pmatrix}
\textbf{1}&0&\textbf{1}\\
0&2&0\\
\textbf{1}&0&-\textbf{1}
\end{pmatrix}~.
\end{equation}
This transforms the 7-vector (\ref{q}) as $\Omega p = q$, i.e.
\begin{equation}
\begin{pmatrix}
y_k\\
t\\
z_k
\end{pmatrix}
\overset{\Omega}{\longrightarrow}
\begin{pmatrix}
\tfrac 12 (y_k + z_k)\\
t\\
\tfrac 12 (y_k - z_k)
\end{pmatrix}
\overset{\text{def}}{=}
\begin{pmatrix}
\lambda_k\\
t\\
x_k
\end{pmatrix}= q~. \qquad (k=1,2,3)
\end{equation}

Cartan's coordinates $y_k$ and $z_k$ and their differential operators have following expressions in terms of the new coordinates $x_k$ and $\lambda_k$,
\begin{equation}
\begin{split}
y_k &= \lambda_k + x_k~, \qquad  \frac {\partial}{\partial y_k} = \frac 12 \left(\frac {\partial}{\partial \lambda_k} + \frac {\partial}{\partial x_k} \right)~,\\
z_k &= \lambda_k - x_k~, \qquad \frac {\partial}{\partial z_k} = \frac 12 \left(\frac {\partial}{\partial \lambda_k} - \frac {\partial}{\partial x_k} \right)~,
\end{split}
\end{equation}
and the invariant quadratic form (\ref{metric}) obtains the form:
\begin{equation} \label{q^2}
p^T\texttt{g}\,p = q^T g \, q = \lambda^2 + t^2 - x^2 ~,
\end{equation}
where
\begin{equation}
g = \Omega^{-1}\texttt{g}\,\Omega^{-1} = {\rm diag}(1,1,1,\underbrace{1,-1,-1,-1}_{\text{Minkowski metric}})
\end{equation}
is the metric tensor in our representation and
\begin{equation}
\lambda^2 = \sum_k \lambda_k\lambda^k~, \qquad x^2 = \sum_k x_kx^k ~.
\end{equation}
We want to associate $t$ and $x_k$ with the coordinates of ordinary Minkowski space-time, while $\lambda_k$ may correspond to some extra time-like dimensions.

Let us also write out Cartan's operators (\ref{CartanTransformations}) in new coordinates,
\begin{equation} \label{CartanTransformations2}
\begin{split}
X_{kk} &= \left(x_k\frac{\partial}{\partial \lambda^k} + \lambda_k\frac{\partial}{\partial x^k}\right) - \frac{1}{3} \sum_i \left(x_i\frac{\partial}{\partial \lambda_i} + \lambda_i\frac{\partial}{\partial x_i}\right)~, \qquad \qquad (i,j,k = 1,2,3)\\
X_{k0} &= \left(\lambda_k \frac{\partial}{\partial t} - t\frac{\partial}{\partial \lambda^k}\right) + \left(x_k\frac{\partial}{\partial t} + t\frac{\partial}{\partial x^k}\right) + \frac{1}{2} \sum_{i,j}\epsilon_{ijk} \left(\lambda^i - x^i\right) \left(\frac{\partial}{\partial \lambda_j} + \frac{\partial}{\partial x_j}\right) ~,\\
X_{0k} &= \left(\lambda_k \frac{\partial}{\partial t} - t\frac{\partial}{\partial \lambda^k}\right) - \left(x_k\frac{\partial}{\partial t} + t\frac{\partial}{\partial x^k}\right) + \frac{1}{2} \sum_{i,j}\epsilon_{ijk} \left(\lambda^i + x^i\right) \left(\frac{\partial}{\partial \lambda_j} - \frac{\partial}{\partial x_j}\right) ~,\\
X_{ij} &= \frac 12 \left(\lambda_i + x_i\right) \left(\frac{\partial}{\partial \lambda^j} + \frac{\partial}{\partial x^j}\right) - \frac 12 \left(\lambda_j - x_j\right) \left(\frac{\partial}{\partial \lambda^i} - \frac{\partial}{\partial x^i}\right)  ~,
\end{split}
\end{equation}
and for the convenience introduce the five classes of $G_2^{NC}$ generators,
\begin{equation}
\label{TFGRPhiDiffGens}
\begin{split}
\Theta_k &= X_{0k} - X_{k0}= - 2\left(x_k\frac{\partial}{\partial t} + t\frac{\partial}{\partial x^k}\right) - \sum_{i,j} \epsilon_{ijk}\left(\lambda^i\frac{\partial}{\partial x_j} - x^j\frac{\partial}{\partial\lambda_i}\right) ~,\\
B_k &= - X_{0k} - X_{k0} = - 2\left(\lambda_k\frac{\partial}{\partial t} + t\frac{\partial}{\partial\lambda^k}\right) - \sum_{i,j} \epsilon_{ijk} \left(\lambda^i\frac{\partial}{\partial\lambda_j} - x^j\frac{\partial}{\partial x_i}\right)~, \\
\Gamma_k &= \sum_{i,j} \lvert \epsilon_{ijk} \rvert X^{ij} = \sum_{i,j} \left|\epsilon_{ijk}\right|\left(x^i\frac{\partial}{\partial\lambda_j} + \lambda^j\frac{\partial}{\partial x_i}\right) ~,  \qquad \qquad (i,j,k = 1,2,3)\\
R_k &= \sum_{i,j} \epsilon_{ijk} X^{ij} = \sum_{i,j} \epsilon_{ijk}\left(\lambda^i\frac{\partial}{\partial\lambda_j} + x^i\frac{\partial}{\partial x_j}\right) ~,\\
\Phi_k &= X_{kk} = \left(x_k\frac{\partial}{\partial\lambda^k} + \lambda_k\frac{\partial}{\partial x^k}\right) - \frac{1}{3}\sum_i \left(x_i\frac{\partial}{\partial\lambda_i} + \lambda_i\frac{\partial}{\partial x_i}\right)~.
\end{split}
\end{equation}

If we denote corresponding group parameters by $\theta_k$, $\beta_k$, $\gamma_k$, $\rho_k$ and $\varphi_k$, then in the new basis the transformations matrix (\ref{Matrix}) can be written as:
\begin{equation} \label{matrixGenerators2}
\begin{pmatrix}
\textbf{B}(\rho) - \textbf{B}(\beta) & -2\beta & \textbf{M}(\gamma,\varphi) - 3\textbf{B}(\theta)\\
-2\beta^T & 0 & 2\theta^T\\
\textbf{M}(\gamma,\varphi) - 3\textbf{B}(\theta) & 2\theta & \textbf{B}(\rho) + \textbf{B}(\beta)
\end{pmatrix} = \theta_k \Theta^k + \beta_k B^k + \gamma_k \Gamma^k + \rho_k R^k + \varphi_k \Phi^k ~.
\end{equation}
In this expression
\begin{equation}
\beta = \begin{pmatrix}
\beta_1 & \beta_2 & \beta_3
\end{pmatrix}^T~, \qquad \qquad
\theta = \begin{pmatrix}
\theta_1 & \theta_2 & \theta_3
\end{pmatrix}^T~,
\end{equation}
are the column vectors, the $3\times 3$ matrices $\textbf{B}$ correspond to the $SO(3)$ group generators, as in (\ref{B}), and $\textbf{M}$ has the form:
\begin{equation}
\textbf{M}(\gamma,\varphi) = \frac{1}{3}\begin{pmatrix}
-2\varphi_1+\varphi_2+\varphi_3 & -3\gamma_3 & -3\gamma_2\\
-3\gamma_3 & \varphi_1-2\varphi_2+\varphi_3 & -3\gamma_1\\
-3\gamma_2 & -3\gamma_1 & \varphi_1+\varphi_2-2\varphi_3
\end{pmatrix}~.
\end{equation}
Then infinitesimal transformations of coordinates of the 7-space (\ref{q^2}) can be written in the form:
\begin{equation} \label{infSmallTransform}
\begin{split}
\lambda'_k &= \lambda_k + \sum_{i,j} \epsilon_{ijk}\left(\beta^i - \rho^i\right)\lambda^j - 2\beta_kt - \sum_{i,j} \left(\epsilon_{ijk}\theta^i + \left|\epsilon_{ijk}\right|\gamma^i\right)x^j - \left(\varphi_k - \frac{1}{3}\sum_i\varphi_i\right) x_k~,\\
t' &= t + 2 \sum_{i} \left(\beta_i\lambda^i + \theta_ix^i\right)~,\\
x'_k &= x_k - \sum_{i,j} \epsilon_{ijk}\left(\beta^i + \rho^i\right)x^j + 2\theta_kt + \sum_{i,j} \left(\epsilon_{ijk}\theta^i - \left|\epsilon_{ijk}\right|\gamma^i\right)\lambda^j - \left(\varphi_k - \frac{1}{3}\sum_i\varphi_i\right) \lambda_k~.
\end{split}
\end{equation}

If we exponentiate the generator matrices (\ref{matrixGenerators2}) separately for each five group parameters, $\rho_k$, $\beta_k$, $\theta_k$, $\varphi_k$ and $\gamma_k$, we will get finite group transformations with either trigonometric or hyperbolic functions. We have two classes of compact rotations by the angles $\rho_k$ and $\beta_k$, 6 rotations in total. Also there are three types of boost-like $G_2^{NC}$-transformations generated by the angles $\theta_k$, $\varphi_k$ and $\gamma_k$, 8 boosts in total because of the constraint (\ref{constraint}), which in our basis is
\begin{equation}
\sum_k \Phi_k = 0~,
\end{equation}
implying only two boost-type $\phi^k$-transformation are linearly independent.
\begin{itemize}
\item{\bf Rotations.} The 3 parameters $\rho_k$ (corresponding to the generators $R_k$), out of 14 of $G_2^{NC}$ transformations, are compact 3-angles that simultaneously rotate the spatial and time-like coordinates $x_i$ and $\lambda_i$ around $k$-th axis ($k\neq i$) within each of these 3-spaces. For example, the finite transformations for $R_1$ are:
\begin{equation}
\begin{split}
\lambda'_1 &= \lambda_1~, \qquad \lambda'_2 = \lambda_2\cos\rho_1 + \lambda_3\sin\rho_1~, \qquad \lambda'_3 = \lambda_3\cos\rho_1 - \lambda_2\sin\rho_1~,\\
t' &= t~,\\
x'_1 &= x_1~, \qquad x'_2 = x_2\cos\rho_1 + x_3\sin\rho_1~, \qquad x'_3 = x_3\cos\rho_1 - x_2\sin\rho_1~.
\end{split}
\end{equation}

The parameters $\beta_k$ (generators $B_k$) also are compact 3-angles that rotate $x_i$ and $\lambda_i$ 3-spaces, but in addition generate simultaneous rotations with the double angle $2\beta_k$ in time-like 4-space of $t$ and $\lambda_k$. For example, the finite transformations for $B_1$ are:
\begin{equation}
\begin{split}
\lambda'_1 &= \lambda_1 \cos \left(2\beta_1\right)- t\sin \left(2\beta_1\right)~, \quad \lambda'_2 = \lambda_2\cos\beta_1 - \lambda_3\sin\beta_1~, \quad \lambda'_3 = \lambda_3\cos\beta_1 + \lambda_2\sin\beta_1~,\\
t' &= t \cos \left(2\beta_1\right) + \lambda_1 \sin \left(2\beta_1\right)~,\\
x'_1 &= x_1~, \qquad x'_2 = x_2\cos\beta_1 + x_3\sin\beta_1~, \qquad x'_3 = x_3\cos\beta_1 - x_2\sin\beta_1~.
\end{split}
\end{equation}

\item{\bf Boosts.} The parameters $\theta_k$ (generators $\Theta_k$) of $G_2^{NC}$ are hyperbolic 3-angles of rotation between $t$ and $x_k$, which in fact are Lorentz boost with the double angle $2\theta_k$. However, these transformations are not pure Lorentz boosts because they also generate hyperbolic rotations by the angles $\theta_k$ between $\lambda_i$ and $x_j$ ($i\neq j \neq k$). For demonstrating this let us write out finite transformations for $\theta_1$:
\begin{equation}
\begin{split}
\lambda'_1 &= \lambda_1~, \qquad \lambda'_2 = \lambda_2\cosh\theta_1 + x_3\sinh\theta_1 ~, \qquad \lambda'_3 = \lambda_3\cosh\theta_1 - x_2\sinh\theta_1~,\\
t' &= t\cosh\left(2\theta_1\right) + x_1\sinh\left(2\theta_1\right)~,\\
x'_1 &= x_1\cosh\left(2\theta_1\right) + t\sinh\left(2\theta_1\right)~, \quad x'_2 = x_2\cosh\theta_1 - \lambda_3\sinh\theta_1~, \quad x'_3 = x_3\cosh\theta_1 + \lambda_2\sinh\theta_1~.
\end{split}
\end{equation}
We get ordinary Lorentz boosts when $x_2 = x_3 = \lambda_2 = \lambda_3=0$.

Another boost-like $G_2^{NC}$-transformations are done by the generators $\Gamma_k$ with the hyperbolic 3-angles $\gamma_k$. For example, the finite transformations for $\Gamma_1$ are:
\begin{equation}
\begin{split}
\lambda'_1 &= \lambda_1~, \qquad \lambda'_2 = \lambda_2\cosh\gamma_1 - x_3\sinh\gamma_1~, \qquad \lambda'_3 = \lambda_3\cosh\gamma_1 - x_2\sinh\gamma_1~,\\
t' &= t ~,\\
x'_1 &= x_1~, \qquad x'_2 = x_2\cosh\gamma_1 - \lambda_3\sinh\gamma_1 ~, \qquad x'_3 = x_3\cosh\gamma_1 - \lambda_2\sinh\gamma_1~.
\end{split}
\end{equation}

The last class of hyperbolic rotations between regular space coordinates $x_k$ and extra time-like coordinates $\lambda_k$ are done by the angles $\varphi_k$. They generate specific diagonal boosts of the spatial coordinates $x_k$ towards the corresponding $\lambda_k$. For example, the finite boost generated by $\Phi_1$ are:
\begin{equation}
\begin{split}
\lambda'_1 &= \lambda_1 \cosh \left(\tfrac 23 \varphi_1 \right) - x_1 \sinh \left(\tfrac 23 \varphi_1\right)~, \\
\lambda'_2 &= \lambda_2 \cosh \left(\tfrac 13 \varphi_1 \right) + x_2 \sinh \left(\tfrac 13 \varphi_1\right)~, \\
\lambda'_3 &= \lambda_3 \cosh \left(\tfrac 13 \varphi_1 \right) + x_3 \sinh \left(\tfrac 13 \varphi_1\right)~,\\
t' &= t~,\\
x'_1 &= x_1 \cosh \left(\tfrac 23 \varphi_1\right) - \lambda_1 \sinh \left(\tfrac 23 \varphi_1\right)~, \\
x'_2 &= x_2 \cosh \left(\tfrac 13 \varphi_1\right) + \lambda_2 \sinh \left(\tfrac 13 \varphi_1\right)~, \\
x'_3 &= x_3 \cosh \left(\tfrac 13 \varphi_1\right) + \lambda_3 \sinh \left(\tfrac 13 \varphi_1\right)~.
\end{split}
\end{equation}
\end{itemize}

We notice that if we assume that $\varphi_k = 0$ and consider only the transformation rules of the Minkowski space-time coordinates, $x_k$ and $t$, then (\ref{infSmallTransform}) will imitate the ordinary infinitesimal Poincar\'{e} transformations,
\begin{equation} \label{Lorentz}
x_k' = x_k - \sum_{i,j}\varepsilon_{kij} \alpha^i x^j + \phi_k t + a_k~, \qquad t' = t  + \sum_i\phi_ix^i + a_0 ~,
\end{equation}
where for the convenience we have introduced the new angles,
\begin{equation}
\alpha^i = \beta^i + \rho^i~, \qquad \phi_k = 2\theta_k~.
\end{equation}
In (\ref{Lorentz}) the parameters for space-time 'translations',
\begin{equation} \label{translations}
a_k = \sum_{i,j} \left(\epsilon_{ijk}\theta^i - \left|\epsilon_{ijk}\right|\gamma^i\right)\lambda^j~, \qquad a_0 = 2\sum_{i}\beta_i \lambda ^i~,
\end{equation}
are represented by the $G_2^{NC}$-rotations into the extra time-like directions $\lambda_k$. So in the language of octonionic geometry \cite{Gogberashvili:2015sga, Gogberashvili:2016ztr, Gogberashvili:2005xb, Gogberashvili:2004na, Gogberashvili:2002wf}, any translation in ordinary space-time is generated by the boosts with $\lambda^k$. Time translations $a_0$ are smooth, since $\beta_k$ are compact angles. However, the angles $\theta_k$ and $\gamma_k$ are hyperbolic and for any spatial translation $a_k$ there exists a horizon (analogue to the Rindler horizon). Also, unlike the translations of Poincar\'e group, the '$G_2^{NC}$-translations' are rotations and thus are non-commutative. For example, commutator between different spatial 'translations' (\ref{translations}) generated by the rotations with the angles $\gamma^i$ produce rotation generators,
\begin{equation}
\left[\Gamma_i,\Gamma_j\right] = -\epsilon_{ijk}R^k~.
\end{equation}
Limits on the parameter of the non-commutativity of coordinates, $[x_i,x_j] < 10^{-8}~Gev^{-2}$ \cite{Chaichian:2000si}, in the theories with non-commutative coordinates \cite{Szabo:2001kg, Gren}, put restrictions on the values of $[a_i, a_j]$, i.e. on the intervals of change of the extra time-like coordinates $\Delta \lambda_k < 10^{-4}~Gev^{-1}$.

Now let us calculate the Casimir operator of $G_2^{NC}$ in our basis and compare it with the Casimirs that are relevant for the Minkowski space-time. Casimir invariants are of primordial importance for a physical model, since they allow us to label the irreducible representations. Eigenvalues of Casimir operators often represent significant dynamical physical quantities, such as angular momentum, elementary particle mass and spin, Hamiltonians of various physical systems etc \cite{Gre-Mul, Sch}.

The rank-2 exceptional Cartan's group $G_2$ has second and sixths order Casimir operators \cite{Bincer:1993jb}. In Cartan's basis (\ref{CartanTransformations}) the second order Casimir of $G_2^{NC}$ has the form:
\begin{equation}
C_2 = 2X_{ij}X^{ij} - \frac{2}{3}\left(X_{k0}X^{0k} + X_{0k}X^{k0}\right)~,
\end{equation}
which in our basis (\ref{TFGRPhiDiffGens}) translates as
\begin{equation} \label{G2DiffCasimir}
\begin{split}
C_2 &= \sum_k\left(\frac{1}{3} \Theta_k^2 - \frac{1}{3}B_k^2 + \Gamma_k^{2\ } - R_k^2 + 2\Phi_k^2\right) = \\
&= 6\left[ t\frac{\partial}{\partial t} + \sum_k \left(\lambda_k\frac{\partial}{\partial\lambda_k} + x_k\frac{\partial}{\partial x_k}\right)\right]+ x^2\frac{\partial^2}{\partial t^2} + \sum_k t^2\frac{\partial^2}{\partial x_k^2} + 2t  \sum_k x_k\frac{\partial^2}{\partial t\partial x_k} + \\
&+ \sum_{i,j,k}\left|\epsilon_{ijk}\right|\left(x_ix_j\frac{\partial^2}{\partial x_i\partial x_j}-x_i^2\frac{\partial^2}{\partial x_j^2}\right) - \lambda^2\left( \frac{\partial^2}{\partial t^2} - \sum_k \frac{\partial^2}{\partial x_k^2}\right) - \sum_k \left(t^2 - x^2\right)\frac{\partial^2}{\partial\lambda_k^2} + \\
&+ 2t \sum_k \lambda_k\frac{\partial^2}{\partial t \partial\lambda_k} + \sum_{i,j}\frac{2}{3}\lambda_ix_j\frac{\partial^2}{\partial\lambda_i\partial x_j} + \sum_{i,j,k}\left|\epsilon_{ijk}\right|\left( \lambda_i\lambda_j\frac{\partial}{\partial\lambda_i\partial\lambda_j}-\lambda_i^2\frac{\partial}{\partial\lambda_j^2}\right)~.
\end{split}
\end{equation}
We note that the first order operator in the first term,
\begin{equation} \label{G2smallInvOp}
c_1 = t\frac{\partial}{\partial t} + \sum_k\left(\lambda_k\frac{\partial}{\partial \lambda_k} + x_k\frac{\partial}{\partial x_k}\right)~,
\end{equation}
itself commutes with all $G_2^{NC}$ generators (\ref{TFGRPhiDiffGens}), therefore putting any other constant coefficient before it in (\ref{G2DiffCasimir}) will not change the Casimir operators property of commuting with all generators. For the convenience we will assume the coefficient in front of the operator (\ref{G2smallInvOp}) in (\ref{G2DiffCasimir}) to be $3$ instead of $6$. Then for the case of the constant extra time-like coordinates, $\lambda_k=const$, i.e. when derivatives with the respect of $\lambda_k$ vanish, we find that (\ref{G2DiffCasimir}) can be expressed as a sum of the second order Casimir operators of the Lorentz group \cite{Tung},
\begin{equation} \label{LorentzCasimirExplicit}
\begin{split}
C_{L} &= 3\left( t\frac{\partial}{\partial t} + \sum_k x_k \frac{\partial}{\partial x_k}\right) + x^2\frac{\partial^2}{\partial t^2} + \sum_k t^2\frac{\partial^2}{\partial x_k^2} +\\
&+  2t  \sum_k x_k\frac{\partial^2}{\partial t\partial x_k} + \sum_{i,j,k}\left|\epsilon_{ijk}\right|\left(x_ix_j\frac{\partial^2}{\partial x_i\partial x_j}-x_i^2\frac{\partial^2}{\partial x_j^2}\right)~,
\end{split}
\end{equation}
and of the Poincar\'e group \cite{Tung},
\begin{equation}
C_P = \frac{\partial^2}{\partial t^2} - \sum_k\frac{\partial^2}{\partial x_k^2}~,
\end{equation}
in the form:
\begin{equation}
\label{LimCasimir}
C_2 = C_{L} -\lambda^2C_P ~.
\end{equation}
This shows that, if one does not consider transformations with extra time-like coordinates $\lambda_k$, the geometrical interpretation of the group $G_2^{NC}$ is consistent with the main properties of the Minkowski space-time.

To conclude, we have considered geometrical applications of Cartan's smallest non-compact exceptional Lie group, $G_2^{NC}$. This group generates specific rotations of the (3+4)-space with the ordinary Minkowski and three extra time-like coordinates. It is shown that in some limiting cases $G_2^{NC}$-rotations imitate standard Poincar\'{e} transformations. In this picture space-time translations are non-commutative and are represented by the rotations towards the extra time-like coordinates. The second order Casimir operator of $G_2^{NC}$ in Minkowski-like basis is found and it is shown that in the case of constant extra dimensions it can be expressed by the Casimir operators of the Lorentz and Poincar\'{e} groups.

\vskip 0.5 cm \noindent
{\bf Acknowledgements:} This work was supported by Shota Rustaveli National Science Foundation of Georgia (SRNSFG) [DI-18-335/New Theoretical Models for Dark Matter Exploration].


\end{document}